\documentclass{article}
\usepackage{epsf}

\begin{document}

\title{Parameters of the Free Core Nutation from VLBI Data%
\footnote{Originally published in Comm. IAA, No.~149, 2003.}}
\author{Z. Malkin, D. Terentev}
\date{Feb 11, 2007}
\maketitle

\begin{abstract}
Several VLBI EOP series were investigated with goal of determination
of parameters of the Free Core Nutation (FCN).
Both the amplitude and period of the FCN were studied using
spectral and wavelet analysis.
Our analysis reveals a variability of both the amplitude
(known also from other investigations) and the period (or phase)
of the FCN nutation.
The FCN amplitude varies in the range about 0.1--0.3 mas,
and the FCN period~--- in the range about 415--490 solar days.
The latter may be also explained by changes in the FCN phase.
Comparison of time variations of the FCN period and amplitude
obtained by different authors and methods shows substantial
discrepancies at the edges of the period of observations.
\end{abstract}

\section{Introduction}

In this paper we investigate variability of the FCN parameters.
Whereas variations of the FCN amplitude was already investigated
(see e.g. \cite{Herring02,Shirai01b}),
variations of the FCN period is not been studied yet.

Modern theory of nutation predicts the steady FCN period of 431.2 sidereal days
\cite{Dehant97}.
The FCN period also have been estimated from VLBI observations, and found
to be about 430--431 sidereal days or about 429--430 solar days
(see, e.g. Table~4 in \cite{Shirai01a}).

In this paper we analyze four VLBI nutation series available
in the IVS data base,
sufficiently long and dense to obtain reliable estimates.
We consider the differences between observed values of nutation
angles and IAU2000A model (which is equivalent to MHB2000 model
without FCN contribution).
For our purpose, we interpret the unpredicted part of observed nutation series
in the FCN frequency band as the FCN contribution.

\section{Data used in analysis}

\begin{figure}
\centerline{\epsfxsize=140mm \epsffile{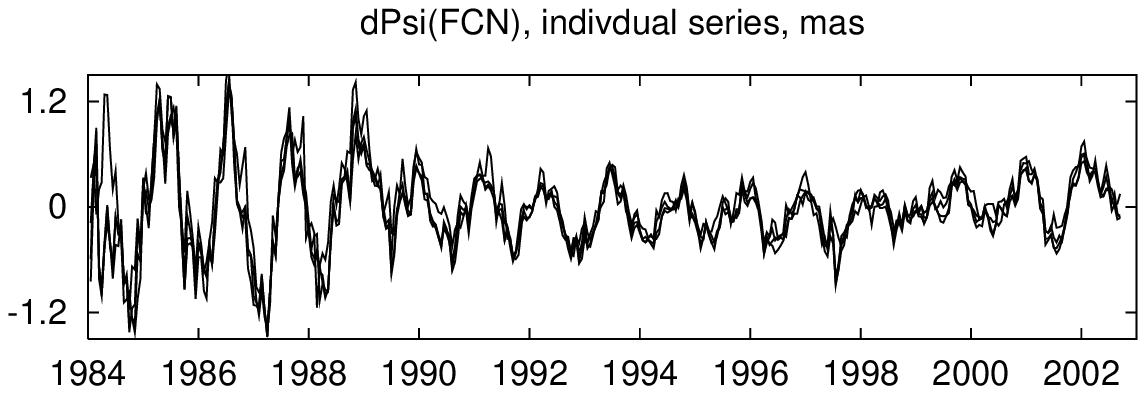}}
\centerline{\epsfxsize=140mm \epsffile{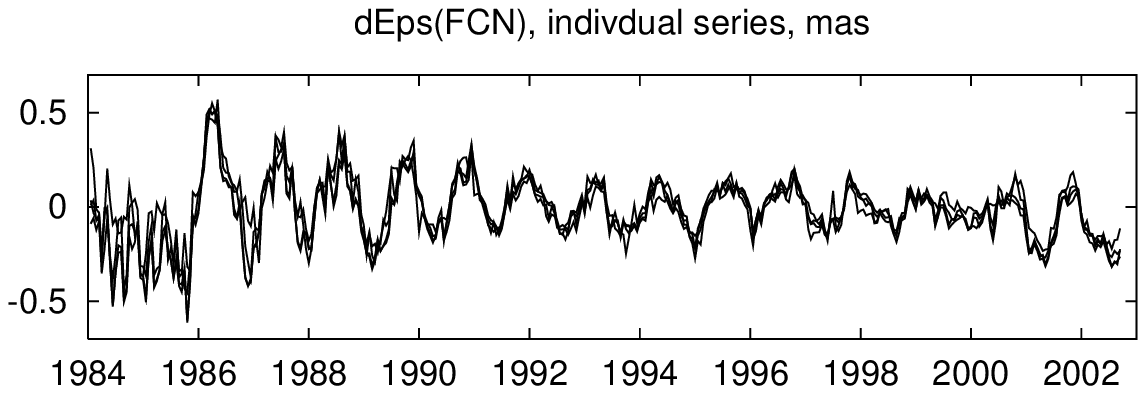}}
\centerline{\epsfxsize=140mm \epsffile{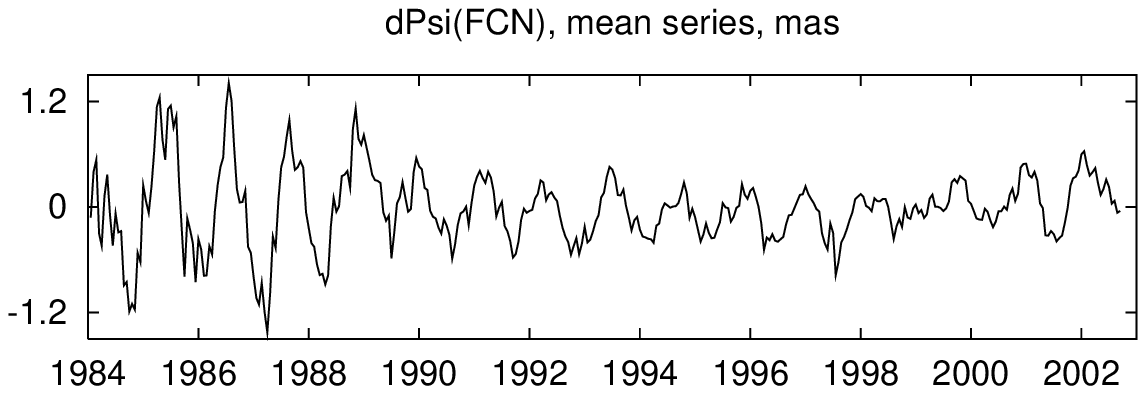}}
\centerline{\epsfxsize=140mm \epsffile{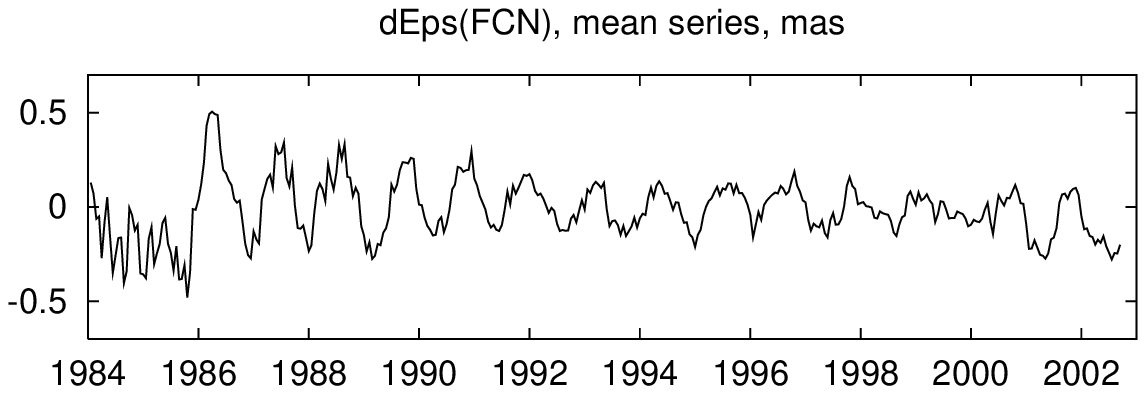}}
\caption{FCN contribution in the individual and mean series.}
\label{fig:series}
\end{figure}

Four celestial pole offset series used in our analysis are
BKG00003, GSF2002C, IAAO0201, USN2002B.
We analyzed both raw (i.e. given on original epochs)
and smoothed (equally spaced by 0.05 year) differences between
observed nutation angles and the IAU2000A model.
For smoothed series we also computed the weighted mean one.
The parameter of smoothing was chosen in such a way to suppress
oscillations with periods less then 1 month.
Common time span for all series is 1984.0--2002.8.
Figure~\ref{fig:series} shows smoothed series used in our analysis.

\section{Analysis and results}

\subsection{Spectral analysis}

\begin{figure}[p]
\centerline{\epsfclipon \epsfxsize=160mm \epsffile{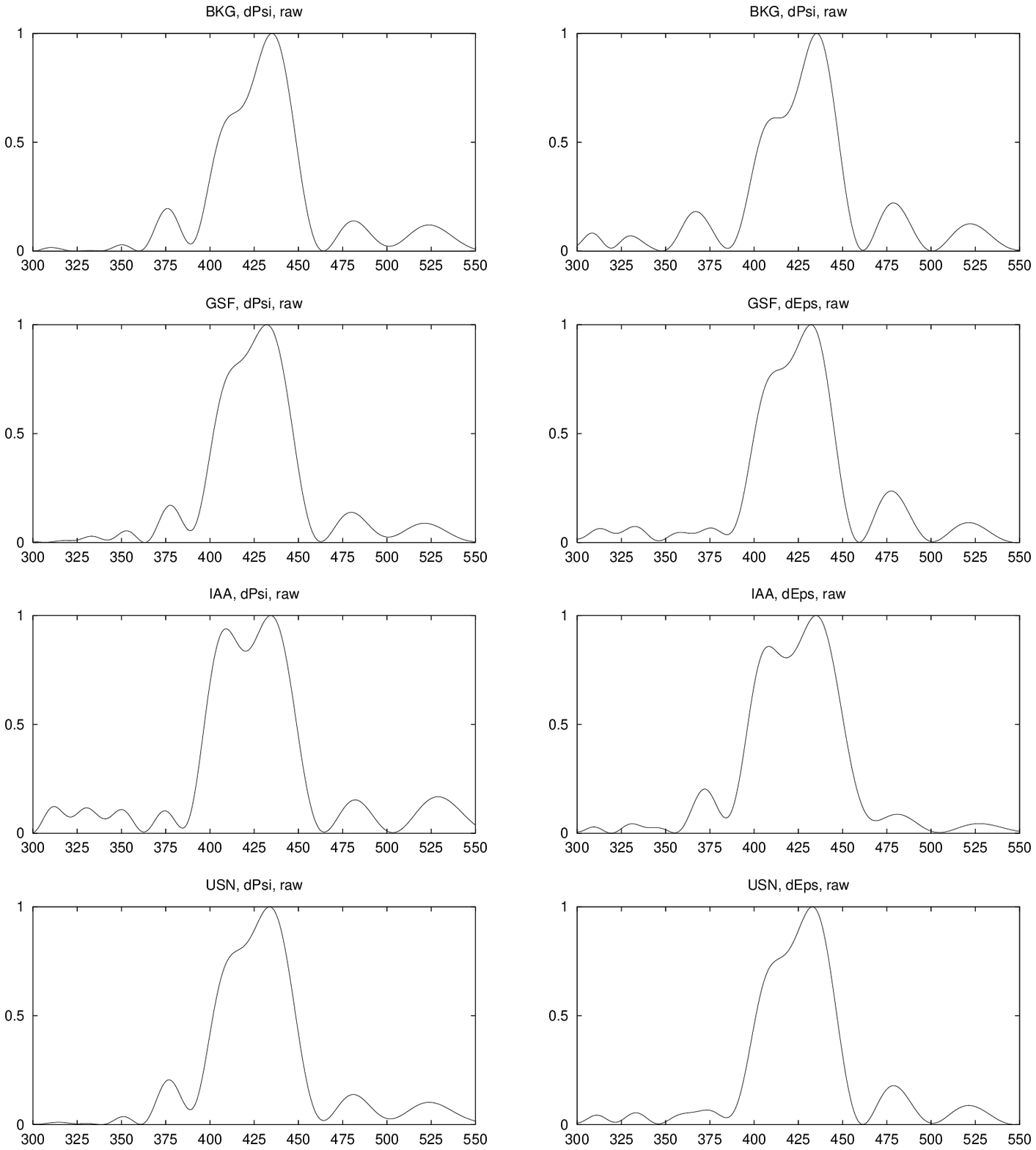}}
\caption{Spectra of raw data, Ferraz-Mello's method, solar days.}
\label{fig:sp_fm_u}
\end{figure}

\begin{figure}[p]
\centerline{\epsfclipon \epsfxsize=160mm \epsffile{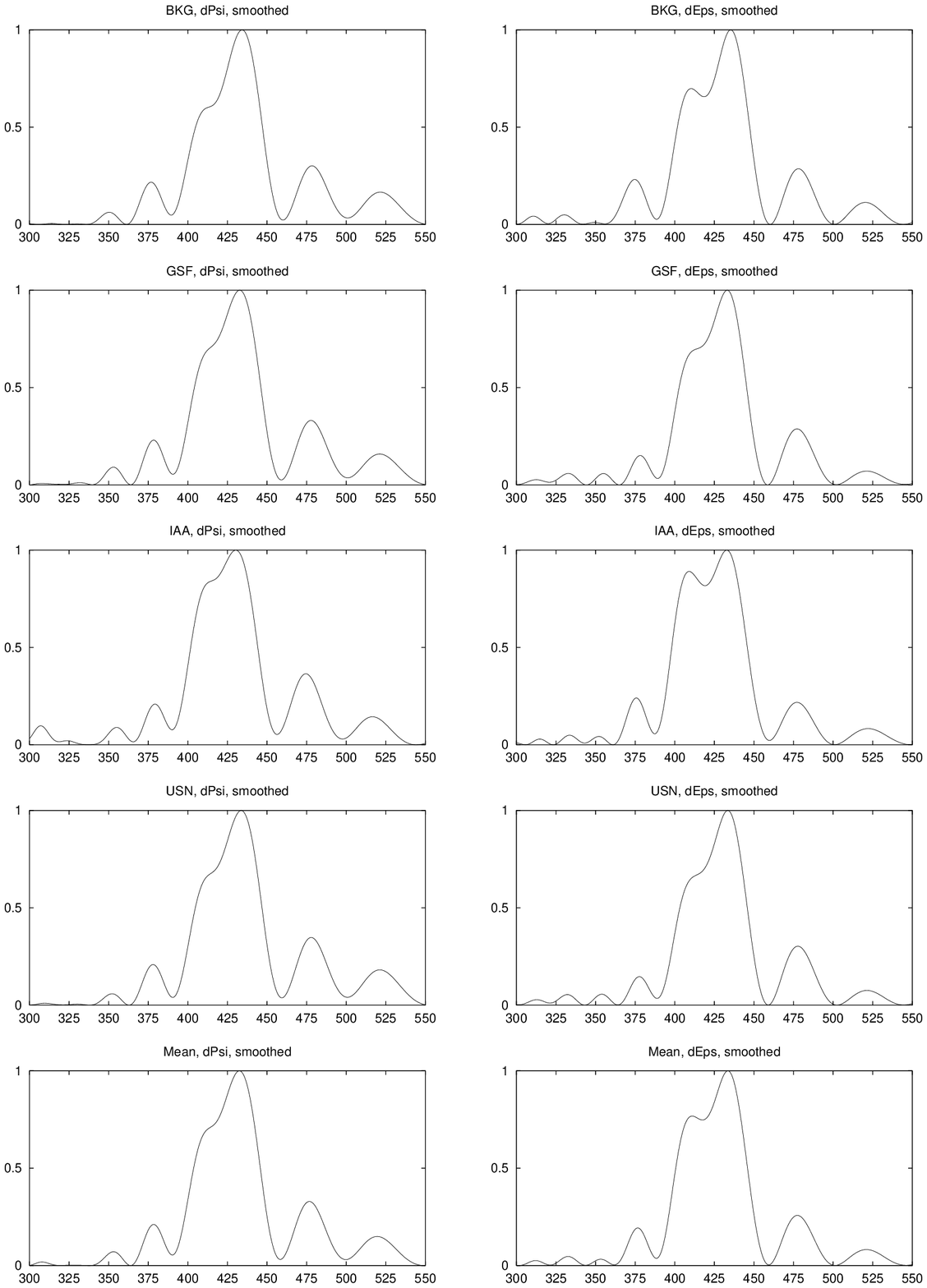}}
\caption{Spectra of smoothed data, Ferraz-Mello's method, solar days.}
\label{fig:sp_fm_e}
\end{figure}

\begin{figure}[p]
\centerline{\epsfclipon \epsfxsize=160mm \epsffile{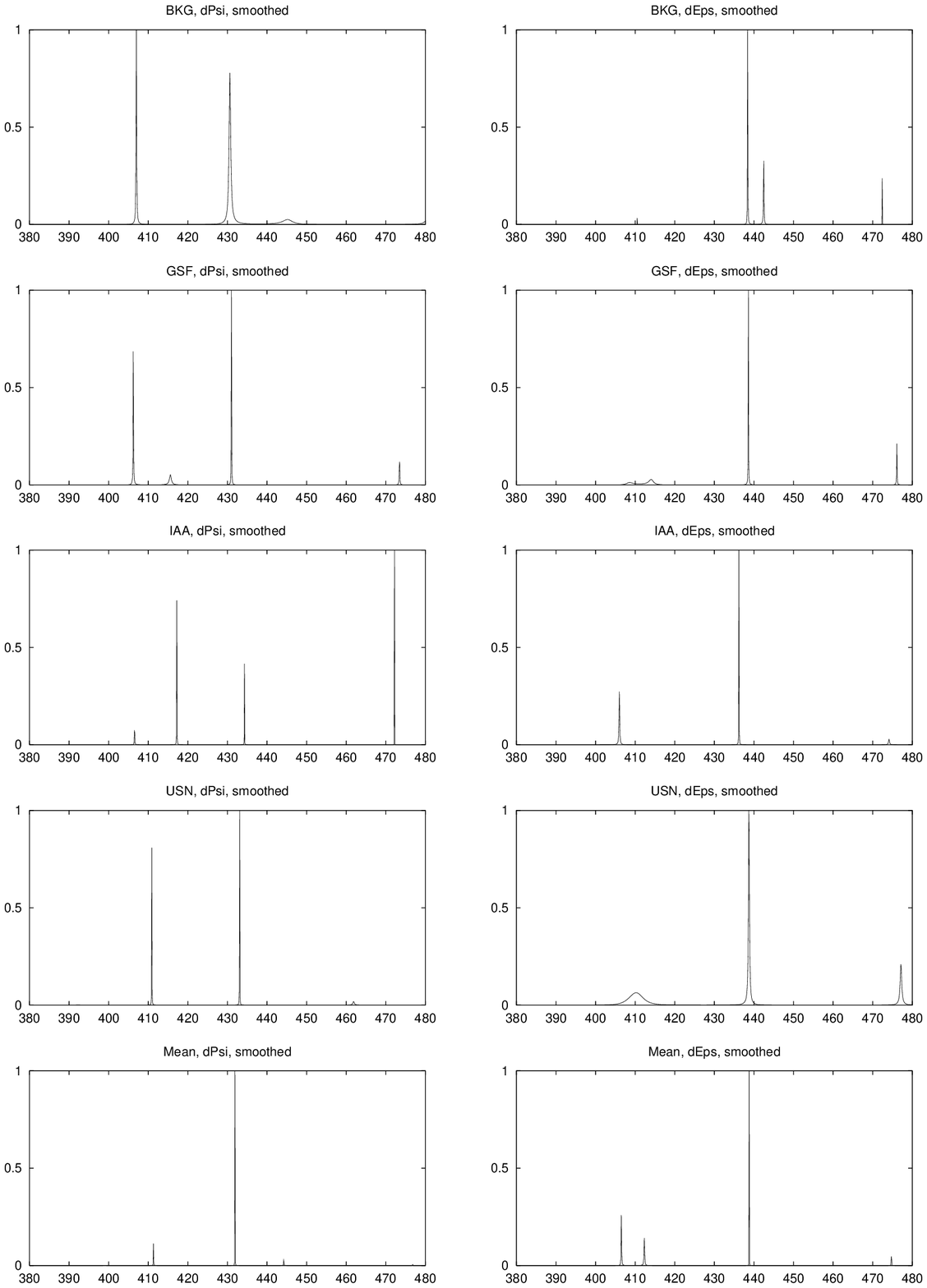}}
\caption{Spectra of smoothed data, Burg's method, solar days.}
\label{fig:sp_burg}
\end{figure}

For estimation of the power spectral density from both raw (unequally
spaced) and smoothed (equally spaced) nutation series
we used the Ferraz-Mello's method \cite{Ferraz-Mello81}
which allows us to process both types of data.
For supplement testing, we also compute the power spectral density using
the Burg's method \cite{Marple87}.
Figures~\ref{fig:sp_fm_u}---\ref{fig:sp_burg} show the normalized
results of spectral estimation, and Table~\ref{tab:periods} presents
the estimates of the FCN period.  Comparison of results shows
reasonable good agreement between the VLBI series, taking into
account that we investigate rather week signal.

\begin{table}
\centering
\smallskip
\caption{Periods of the FCN contribution, solar days.}
\label{tab:periods}
\tabcolsep=7.5pt
\smallskip
\begin{tabular}{llccccc}
\hline
\multicolumn{7}{c}{~} \\[-1ex]
Series   & Method          &  BKG  &  GSF  &  IAA  &  USN  & Mean   \\
\multicolumn{7}{c}{~} \\[-1ex]
\hline
\multicolumn{7}{c}{~} \\[-1ex]
\multicolumn{7}{c}{$\Delta\psi$} \\
\multicolumn{7}{c}{~} \\[-1ex]
Raw      & Ferraz-Mello    & 435.0 & 432.2 & 434.4 & 433.7 & ---   \\
Smoothed & Ferraz-Mello    & 434.2 & 432.7 & 430.3 & 433.7 & 432.5 \\
Smoothed & Burg            & 430.6 & 431.0 & 434.3 & 433.1 & 431.9 \\
Smoothed & Burg (--2000.2) & 433.4 & 430.0 & 428.5 & 433.4 & 431.9 \\
\multicolumn{7}{c}{~} \\[-1ex]
\multicolumn{7}{c}{$\Delta\varepsilon$} \\
\multicolumn{7}{c}{~} \\[-1ex]
Raw      & Ferraz-Mello    & 435.4 & 432.2 & 435.0 & 432.9 & ---   \\
Smoothed & Ferraz-Mello    & 435.4 & 433.1 & 432.9 & 433.5 & 433.5 \\
Smoothed & Burg            & 438.4 & 438.6 & 436.2 & 438.7 & 438.8 \\
Smoothed & Burg (--2000.2) & 431.7 & 428.5 & 429.5 & 430.1 & 429.9 \\
\multicolumn{7}{c}{~} \\[-1ex]
\multicolumn{7}{c}{Mean of $\Delta\psi$ and $\Delta\varepsilon$} \\
\multicolumn{7}{c}{~} \\[-1ex]
Raw      & Ferraz-Mello    & 435.2 & 432.2 & 434.7 & 433.3 & ---   \\
Smoothed & Ferraz-Mello    & 434.8 & 432.9 & 431.6 & 433.6 & 433.0 \\
Smoothed & Burg            & 434.5 & 434.8 & 435.2 & 435.9 & 435.4 \\
Smoothed & Burg (--2000.2) & 432.6 & 429.2 & 429.0 & 431.8 & 430.9 \\
\multicolumn{7}{c}{~} \\[-1ex]
\hline
\smallskip
\end{tabular}
\end{table}

The average estimated value of the FCN period is of about 434 solar days
(about 435 sidereal days).  This value is substantially greater than one
found in \cite{Shirai01a} (431.0$\pm$0.6 sidereal days).
However, when we used for spectral analysis
only nutation series cut at the epoch 2000.2 which corresponds
to the data span used in \cite{Shirai01a},
we obtain the FCN period of about 432 sidereal
days which is close to found in \cite{Shirai01a}
(see the last line in each section of Table~\ref{tab:periods}).

It is interesting to note a second period in the investigated
frequency band, about 410 days.  This period can be clearly
seen in the Burg spectra, and as well it is discernible in
the Ferraz-Mello spectra.  Evidently a term with this period
deserves further investigation.\footnote{After this paper was
published in its original form, Krasinsky\& Vasilyev
(CMDA, 2006, v.~96, 219) gave a theoretical explanation of
this FCN mode.}

\subsection{Wavelet analysis}

\begin{figure}
\centerline{\epsfclipon \epsfxsize=160mm \epsffile{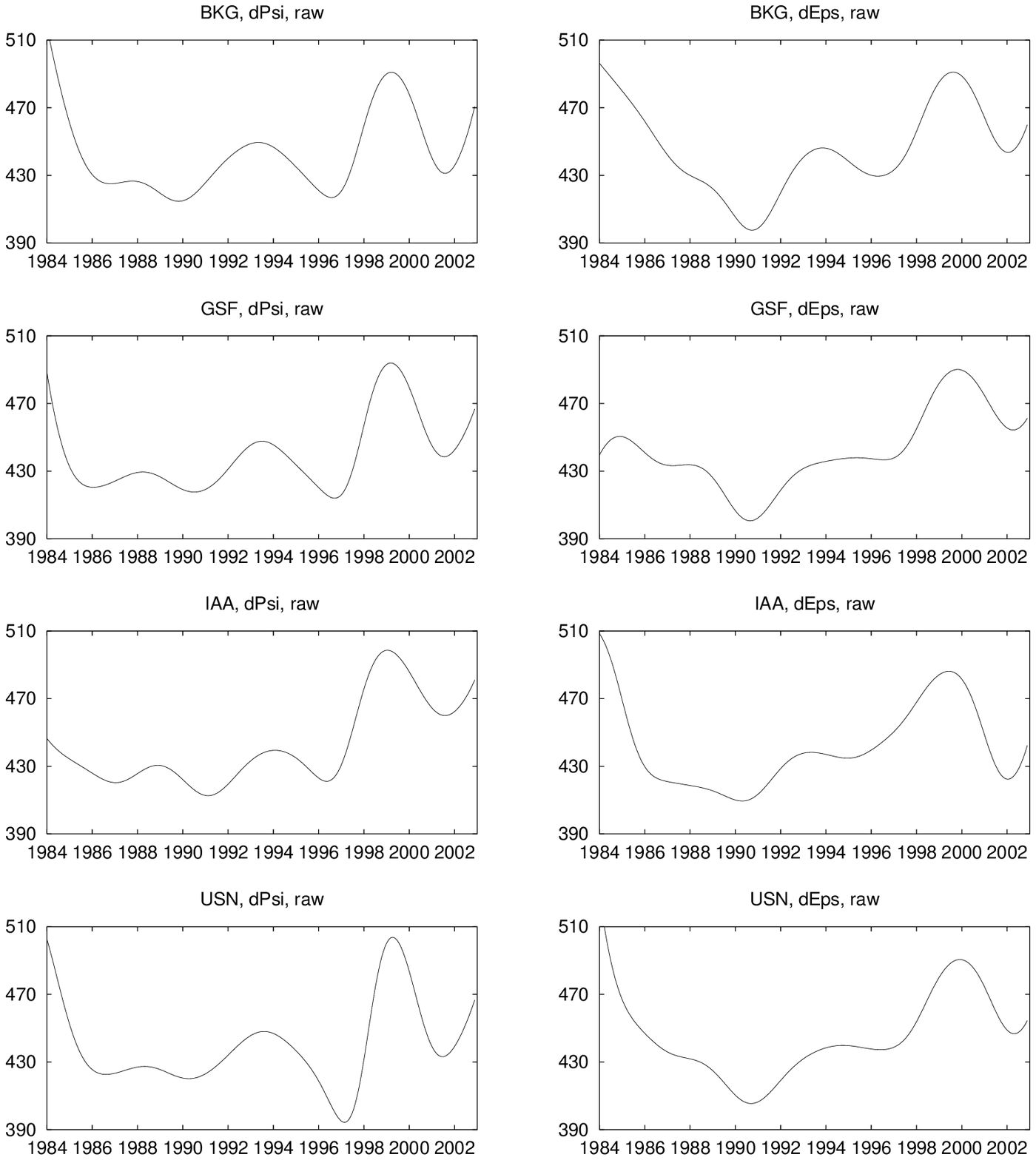}}
\caption{Variations of the FCN period with time, raw data, solar days.}
\label{fig:wwz_up}
\end{figure}

\begin{figure}
\centerline{\epsfclipon \epsfxsize=160mm \epsffile{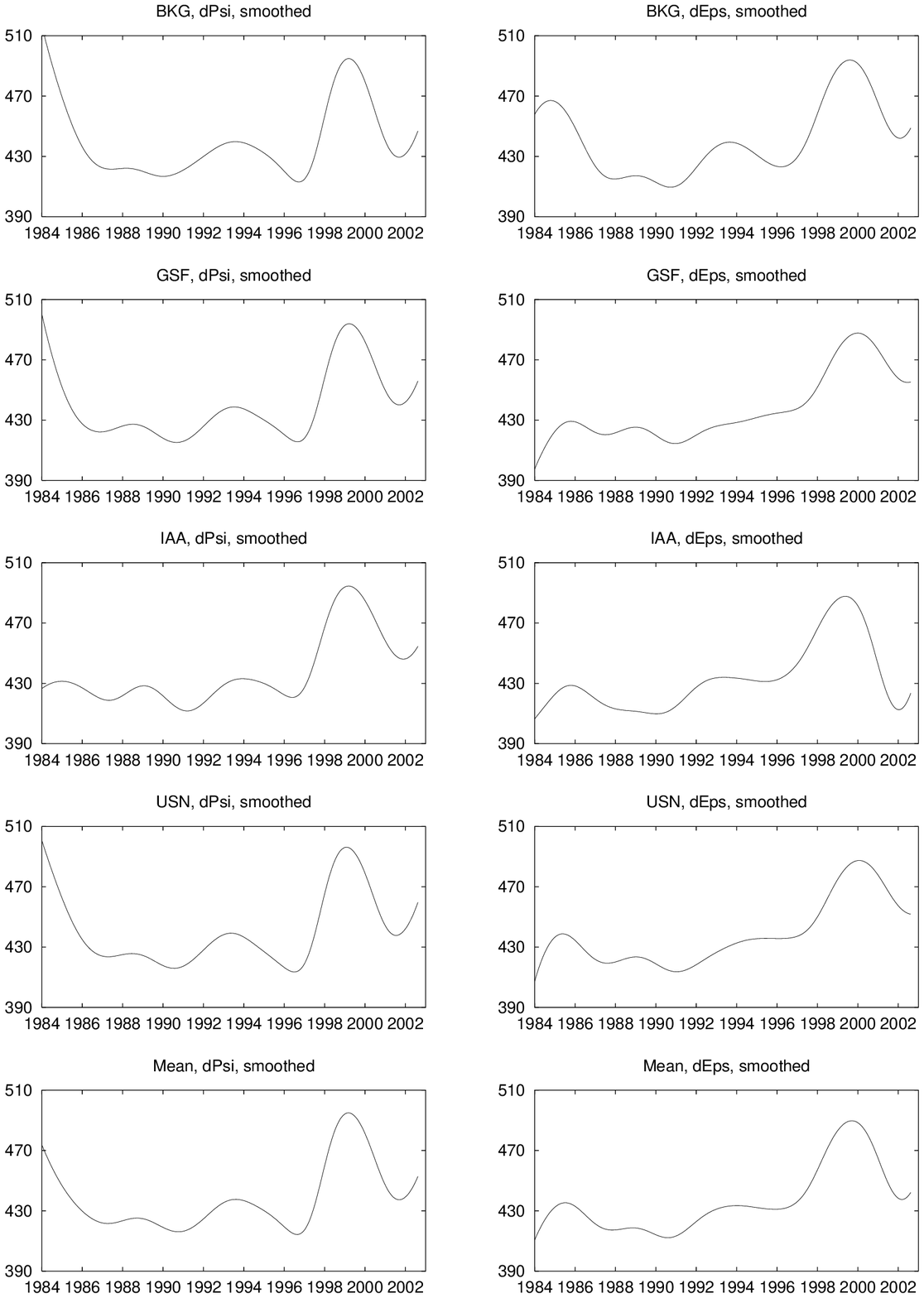}}
\caption{Variations of the FCN period with time, smoothed data, solar days.}
\label{fig:wwz_ep}
\end{figure}

\begin{figure}
\centerline{\epsfclipon \epsfxsize=160mm \epsffile{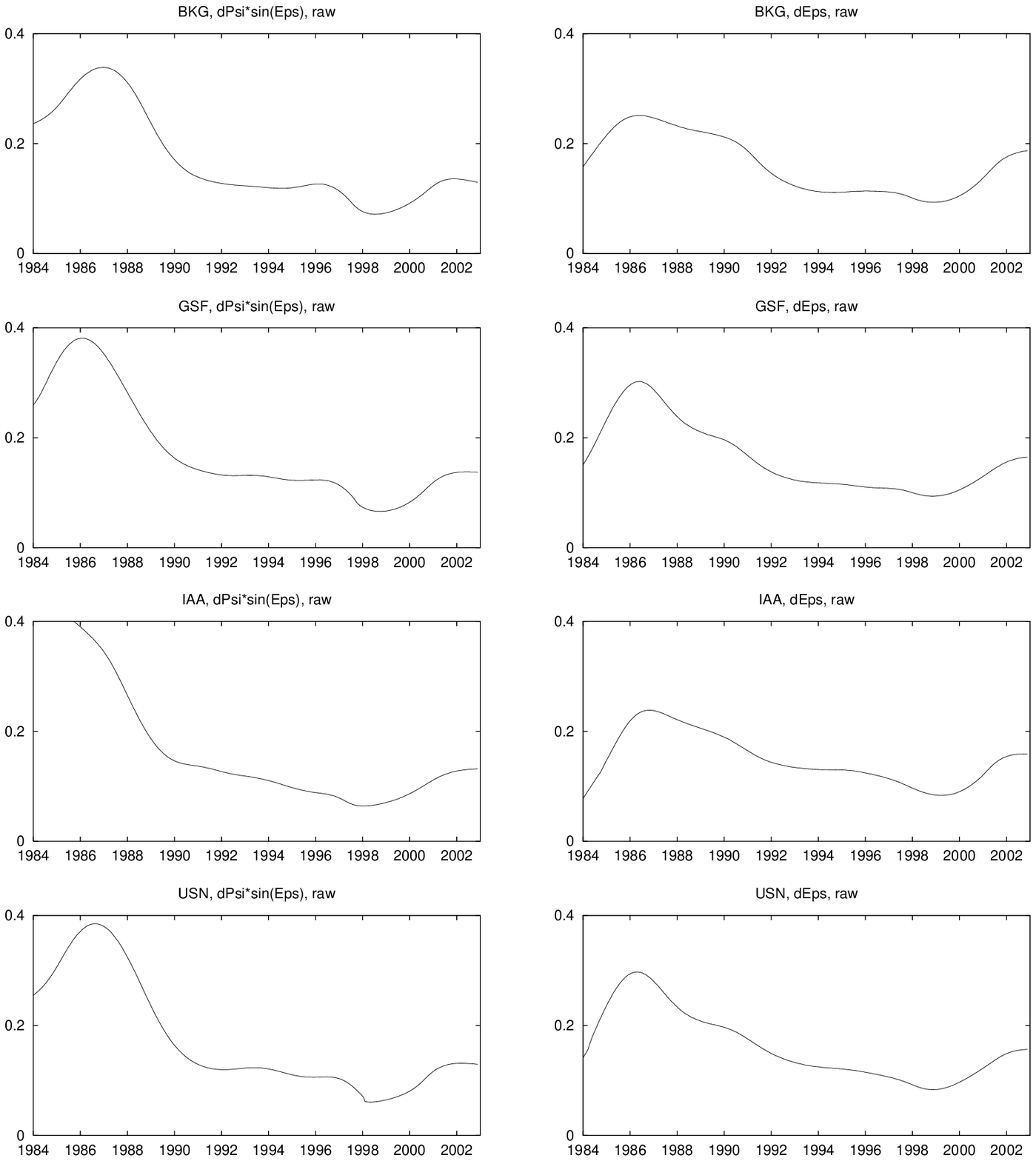}}
\caption{Variations of the FCN amplitude with time, raw data.}
\label{fig:wwz_ua}
\end{figure}

\begin{figure}
\centerline{\epsfclipon \epsfxsize=160mm \epsffile{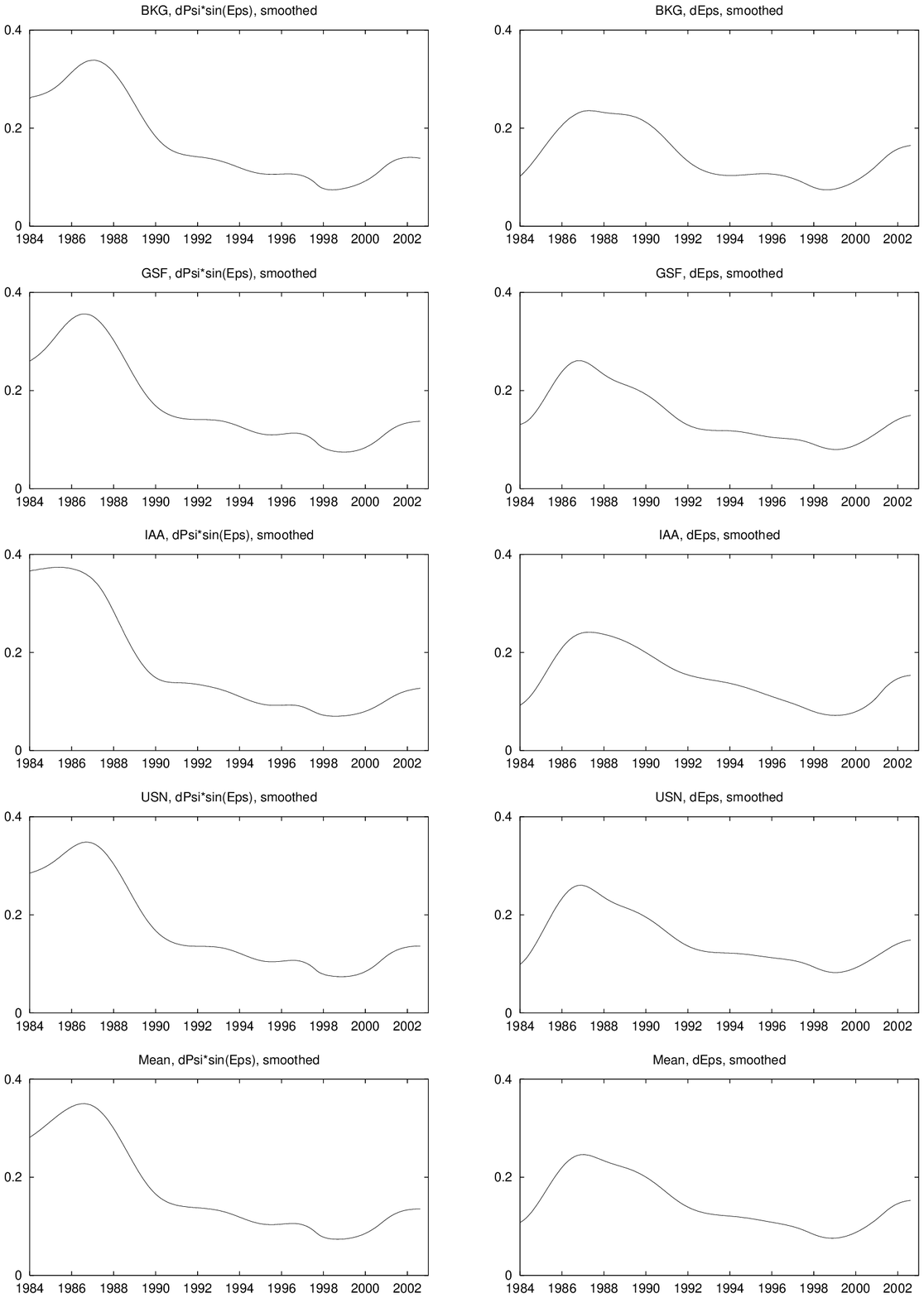}}
\caption{Variations of the FCN amplitude with time, smoothed data.}
\label{fig:wwz_ea}
\end{figure}

At the next step we had applied the wavelet analysis to all the nutation series
to investigate the time variations of the FCN period (phase) and amplitude,
which is the main goal of our study\footnote{After this paper was published
in its original form, the authors were pointed out that perhaps the first
attempt to apply the wavelet analysis to investigation of the FCN was made
by Schmidt \& Schuh (ZfV, 1999, No.~1, 24).}.
For this analysis we used program WWZ, developed by the American
Association of Variable Star Observers and available as executable at the
http://www.aavso.org/cdata/wwz.shtml.
Theoretical background of this method can be found in~\cite{Foster96c}.
The results of the wavelet analysis are presented in
Figures~\ref{fig:wwz_up}--\ref{fig:wwz_ea}.

\begin{figure}
\centerline{\epsfclipon \epsfxsize=160mm \epsffile{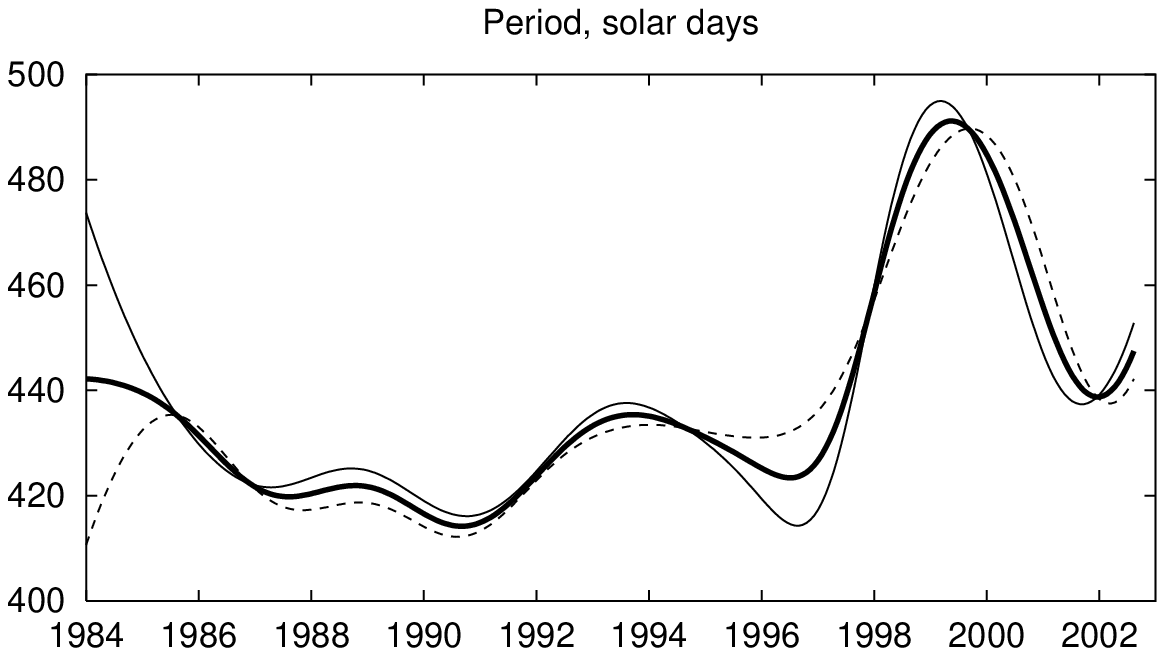}}
\centerline{\epsfclipon \epsfxsize=160mm \epsffile{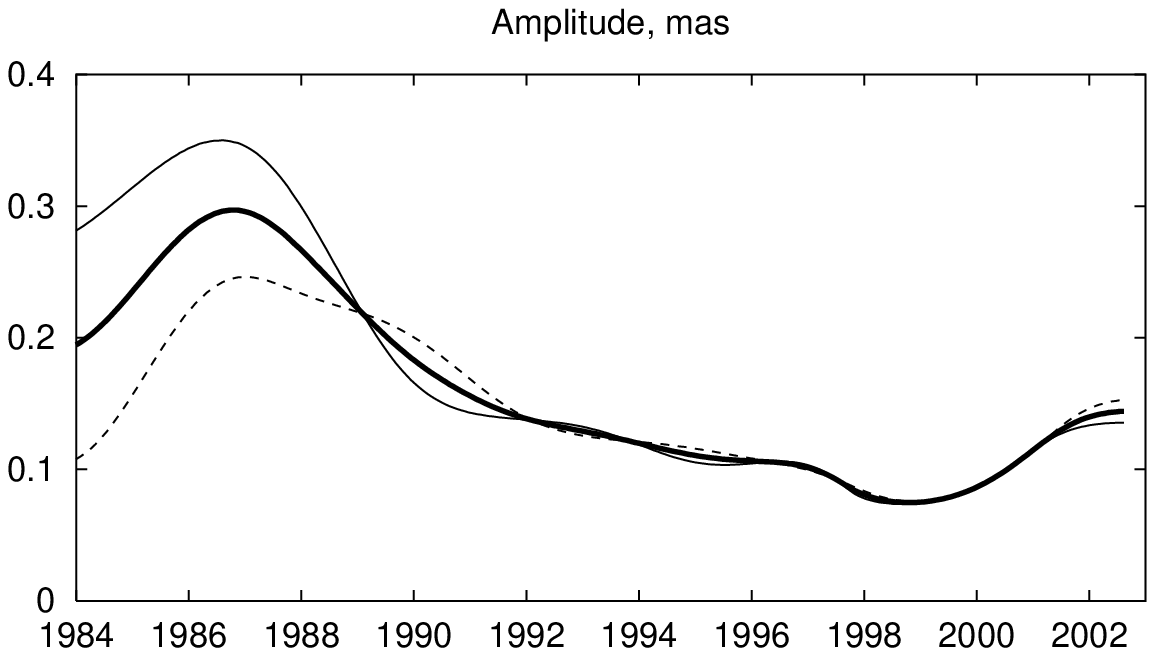}}
\caption{Variations of the FCN period and amplitude with time;
 $\Delta\psi\ast\sin(\varepsilon)$ (solid line),
 $\Delta\varepsilon$ (dashed line), and mean of
 $\Delta\psi$ and $\Delta\varepsilon$ (bold line).}
\label{fig:wwz_mean}
\end{figure}

Figure~\ref{fig:wwz_mean} presents the final results of the present
investigation.
It should be mentioned that based on the comparison of FCN amplitudes
found here and previous investigations \cite{Malkin02g},
we consider the results obtained before 1990 seems to be not very reliable.

Of course, an important question arising from the obtained result
is whether the variations of the period found from our analysis is an actual
geophysical signal or an artifact caused by inadequate computational
procedures.
One can see that large increasing of the FCN period after $\approx$1998
corresponds to relatively low amplitude of the FCN oscillation.
We have performed some tests to estimate how result of wavelet analysis
depends on variable amplitude of input signal.

For test purposes we used several artificial signals, and also we
constructed new series as original one normalized by
found variations of the FCN amplitude.  The latter provides the
FCN contribution series with near-unity amplitude with all
other peculiarities inherited from real nutation series.
The variations of the period found for the test series are
practically the same as for real data.

After all test, our conclusion is that found variations of the FCN period
cannot be explained by computational errors.
Besides, the results of spectral analysis made for different subset
of data also corroborate our conclusion.

\begin{figure}[p]
\centerline{\epsfclipon \epsfxsize=160mm \epsffile{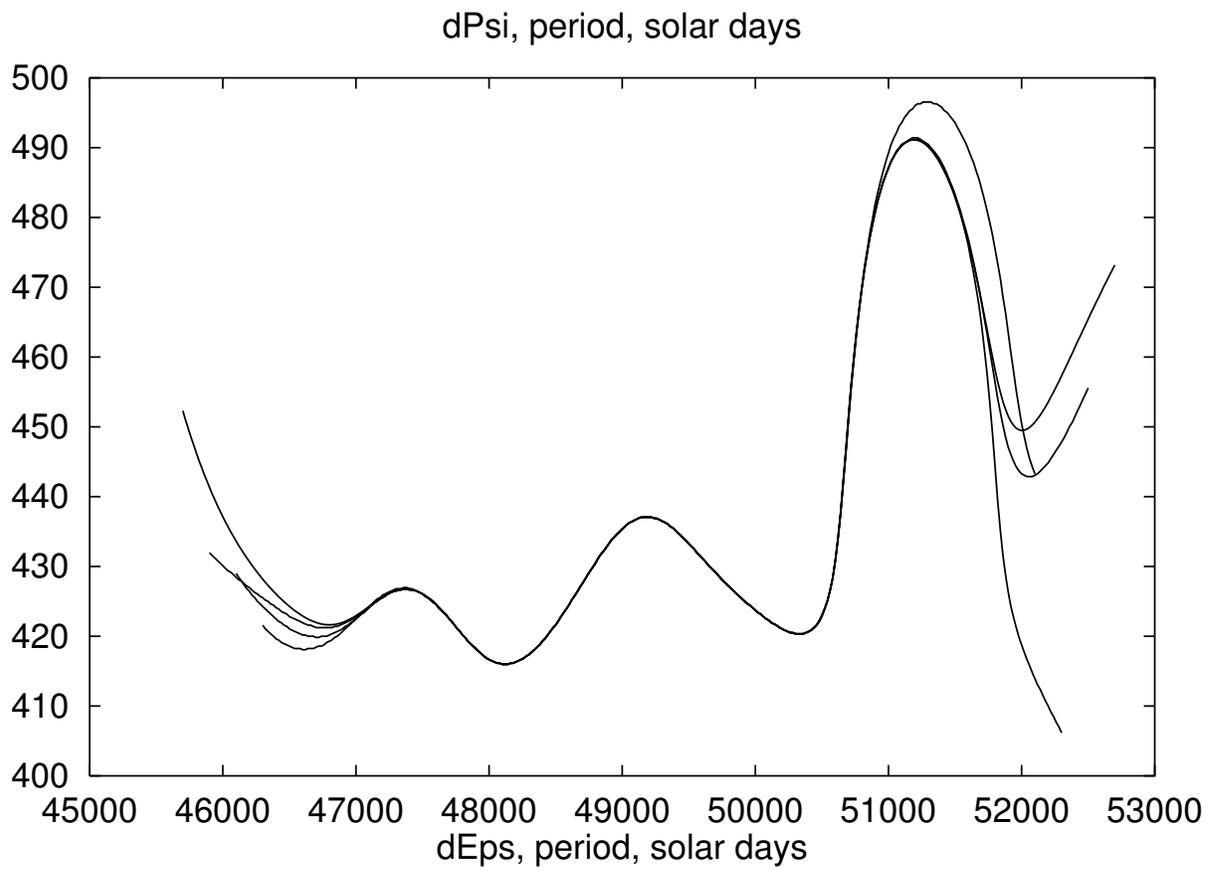}}
\centerline{\epsfclipon \epsfxsize=160mm \epsffile{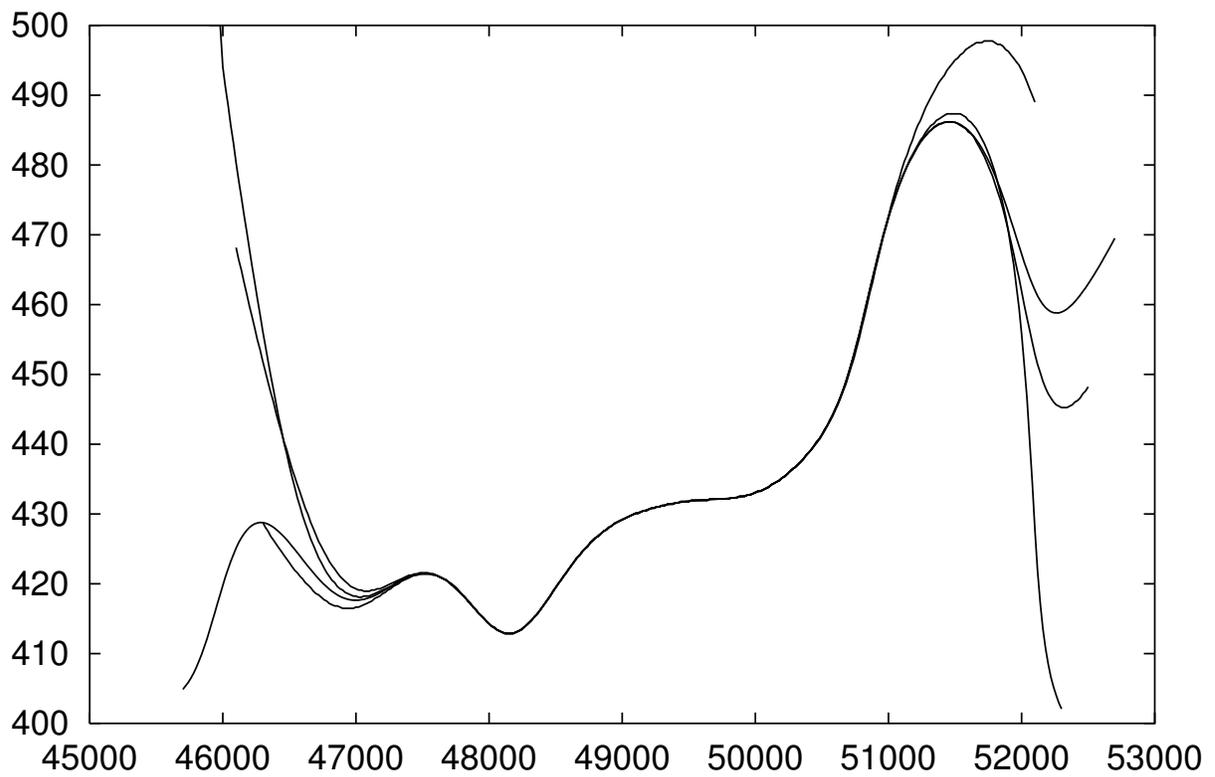}}
\caption{The edge effect in the wavelet analysis, the FCN period.}
\label{fig:wwz_edge_p}
\end{figure}

\begin{figure}[p]
\centerline{\epsfclipon \epsfxsize=160mm \epsffile{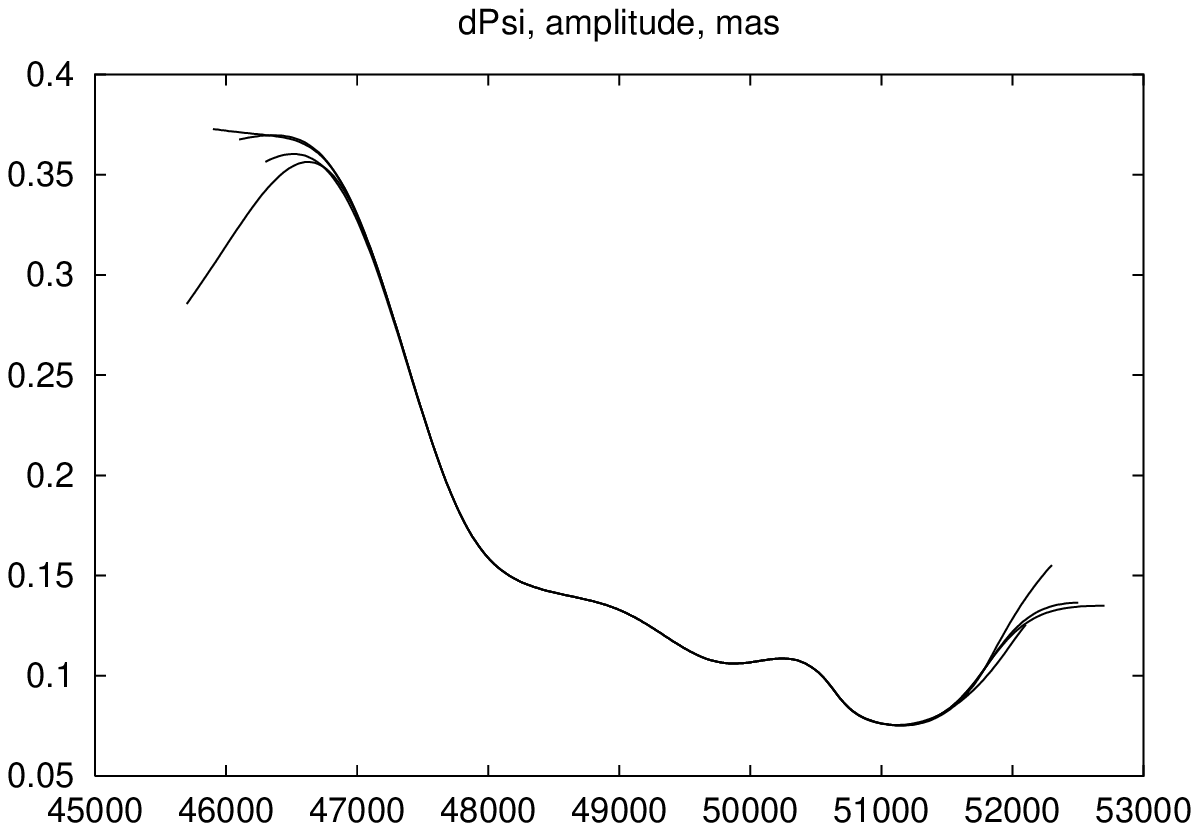}}
\centerline{\epsfclipon \epsfxsize=160mm \epsffile{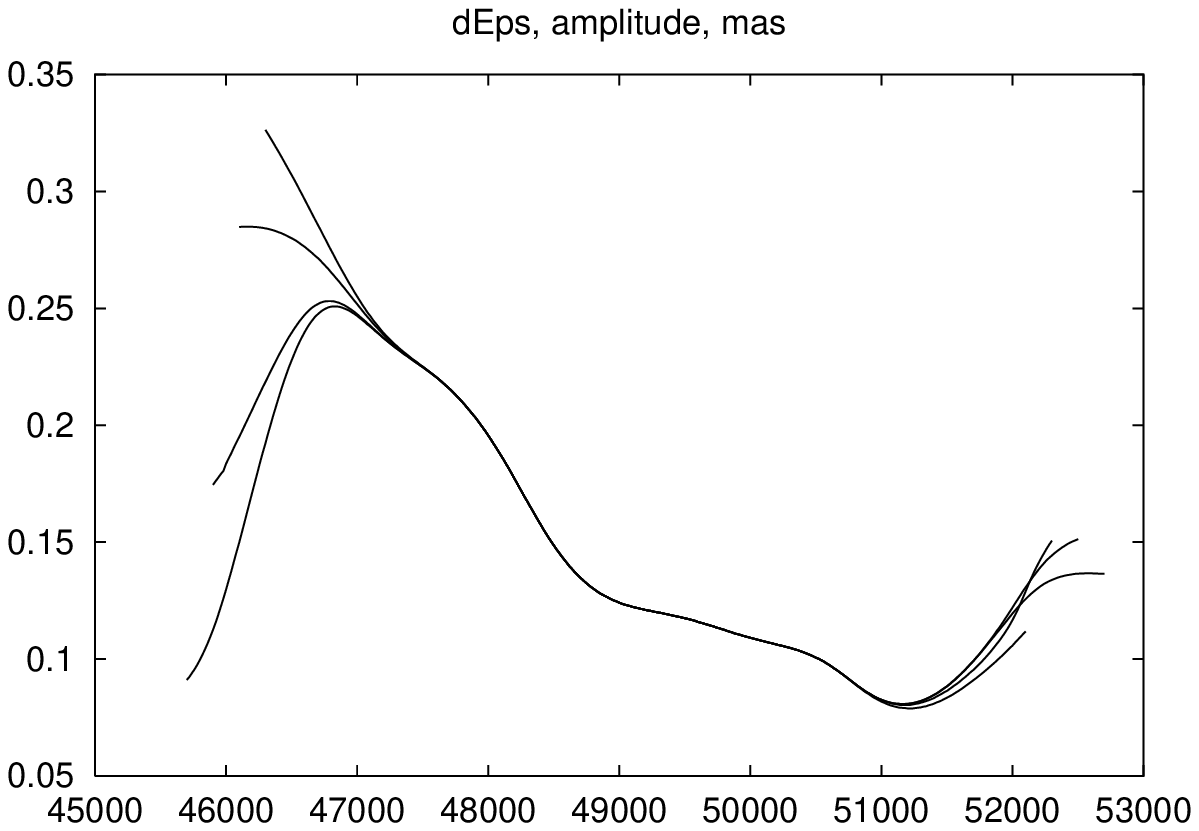}}
\caption{The edge effect in the wavelet analysis, the FCN amplitude.}
\label{fig:wwz_edge_a}
\end{figure}

One of the important points to be investigated is the edge effect
which may lead to misinterpreting of the results of the wavelet
analysis for the first and the last epochs.
For this purpose we performed a special test with WWZ.
We computed the FCN period and amplitude variations for several
series starting with original one and cutting the first and the
last 200 points from it. Then this process was repeated and in that way
four test series were obtained, each starts 200 days later and ends
200 days earlier than previous one.
Figures \ref{fig:wwz_edge_p} and \ref{fig:wwz_edge_a} show the result
of this test.

From this test, we can conclude that the edge effect may affect the
result at the first and the last 500--1000 days of the interval
under investigation.
Evidently, this fact should be accounted for to determinate the
start point for prediction of the FCN contribution to nutation.

\section{Discussion and conclusions}

\begin{figure}
\centerline{\epsfclipon \epsfxsize=160mm \epsffile{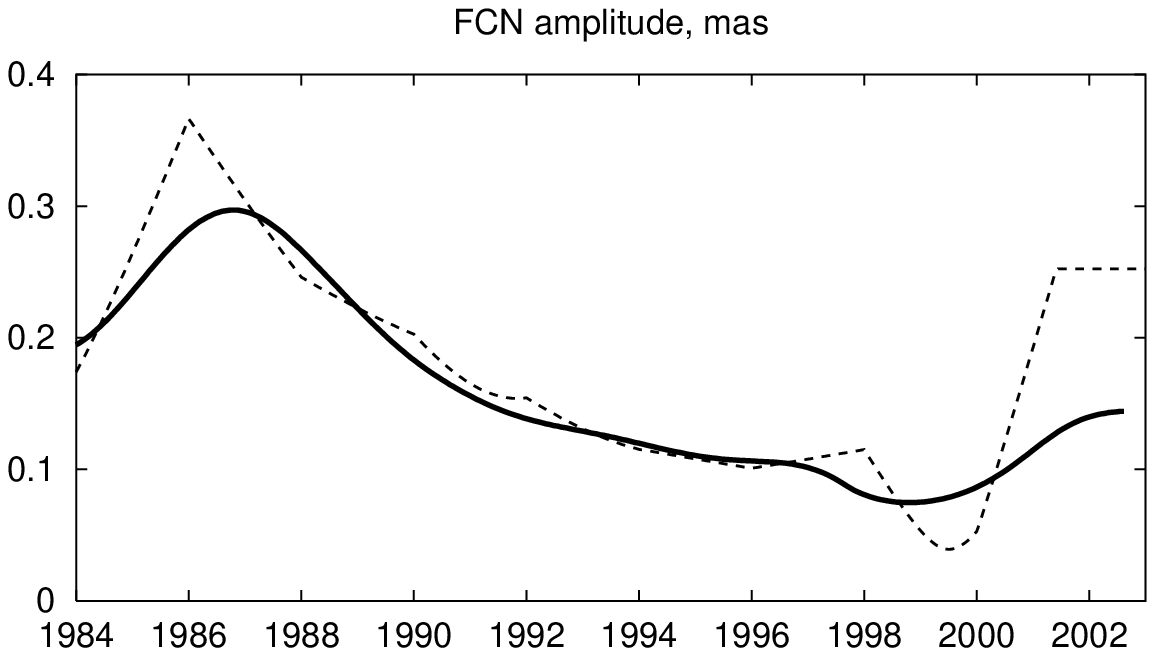}}
\caption{Variations of the FCN amplitude with time found in the
present study (solid line) and implemented in the MHB2000 model
(dashed line).}
\label{fig:ampcomp}
\centerline{\epsfclipon \epsfxsize=160mm \epsffile{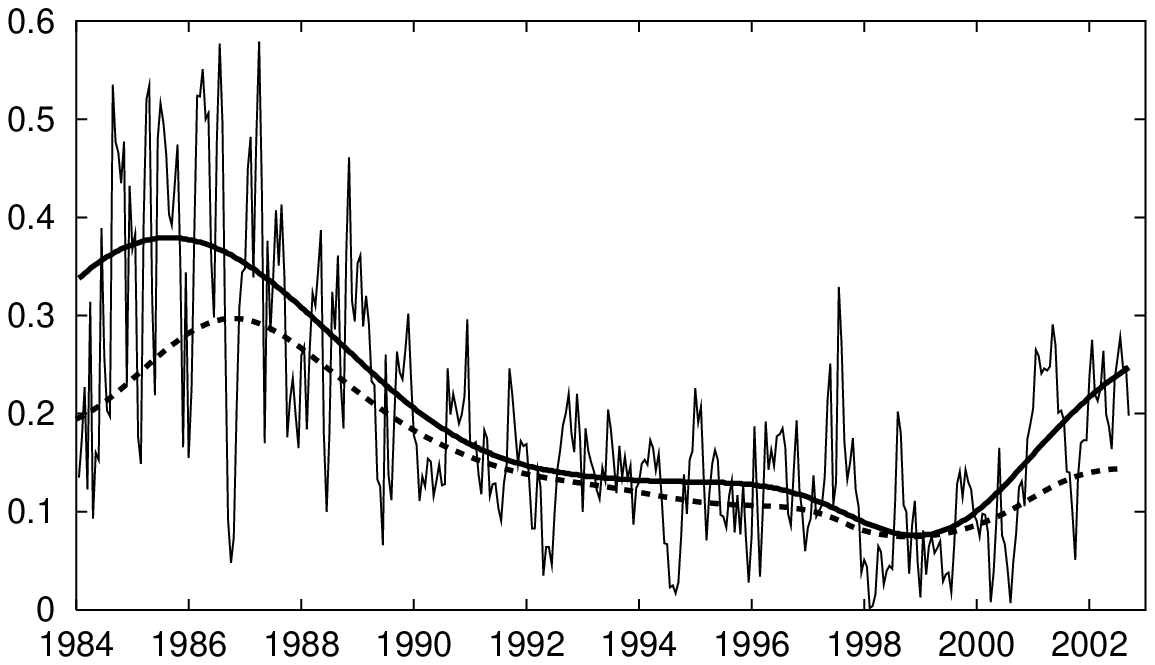}}
\caption{Variations of the FCN amplitude with time:
solid line~--- original amplitudes, bold line~--- smoothed amplitudes,
bold dashed line~--- amplitude variations found from the wavelet analysis.}
\label{fig:ampl_ma}
\end{figure}

Variations of the FCN amplitude found in this investigation
are close to ones used in the MHB2000 model, except the edge
intervals (see Figure~\ref{fig:ampcomp}).  A reasons of these
discrepancies may be insufficient quality of the VLBI data in the
earlier 1980th and the edge effect present
in the wavelet analysis results, as discussed above.

As a supplement test, we computed variations of the FCN amplitudes
immediately from the VLBI series.  For this purpose we used the
smoothed differences between observed nutation angles and model
described above.

Taking into account that the FCN contributions
in $\Delta\psi$ and $\Delta\varepsilon$ are two projections of the same variations in
the Earth rotation velocity, we can compute the FCN amplitude as
$Amp(FCN) = \sqrt{d(\Delta\psi)^2+d(\Delta\varepsilon)^2}$.
Figure~\ref{fig:ampl_ma} shows the result of comparison.
Comparing Figures \ref{fig:ampcomp} and \ref{fig:ampl_ma}
one can see that the direct computation of the FCN amplitude
shows better agreement with the MHB2000 model.
We can expect that amplitude estimates obtained with WWZ method
are not good enough at the edges of the time interval (see also
the discussion of the edge effect in wavelet analysis above).

\begin{figure}
\centerline{\epsfclipon \epsfxsize=160mm \epsffile{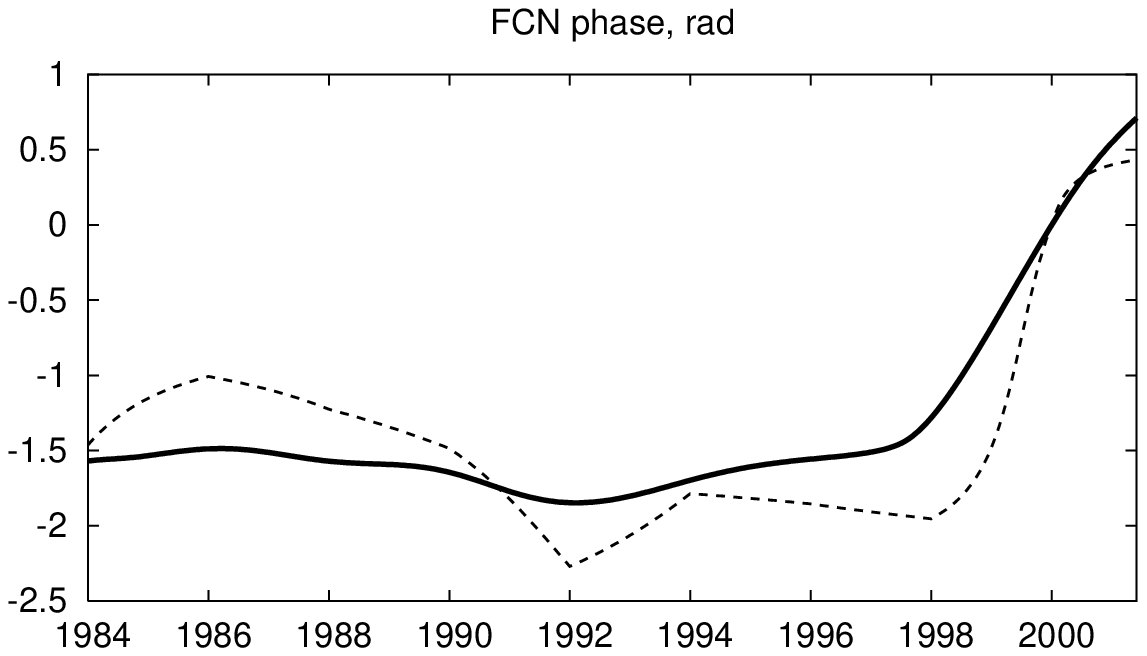}}
\caption{Variations of the FCN phase with time found in the
present study (solid line) and computed from the MHB2000 model
(dashed line).}
\label{fig:phasecomp}
\end{figure}

Comparison of the FCN phase variations found in this study
and computed from the amplitudes of sine and cosine FCN terms
of the MHB2000 model is presented in Figure~\ref{fig:phasecomp},
after removing the linear phase change corresponding to the
FCN with permanent period.
One can see that the FCN phase variations are similar in two approaches,
though ours provides more smooth variations, and so for the FCN period
variations.

The FCN period (phase) most likely varies with time.
Probably, change in the period is physically connected with change in amplitude.
On the other hand, one can see that the variations of the FCN period
show clear periodicity with a period about 5 years, whereas variations
of the FCN amplitude does not show such an effect.

Another reason of the observed behavior of the FCN period maybe
variability of the FCN phase.
Analogous effect was found also at the Chandler frequency \cite{Vondrak88},
for which dependence of the period on amplitude, and the phase jump
occurred during the period of the lowest amplitude were also found.

It is interesting, that the Chandler wobble period also
decreased in $\approx$1986--1988, and
increased in $\approx$1989--1996 (see \cite{Hopfner03,Schuh01}).
Unfortunately, Polar Motion series studied in those papers are much
shorter than one analyzed here to perform a reliable comparison.

Variations of FCN amplitudes show several possible epochs of the excitation
of the FCN, most of them are close to ones detected in \cite{Shirai01b}.

Some tests we performed allow us to make a conclusion that
investigated nutation series really contain such a signal with variable
amplitude and period (phase).
However, as stated above, we interpret the differences between observed
nutation and the IAU2000A model as the FCN contribution, which may be too
strong assumption.  Possible interference of the FCN and other nutation
frequencies should be carefully investigated.  In particular, the authors
of \cite{Herring02} pointed out a possible interference with near-yearly
nutation terms, but the investigated period of observations seems to be
long enough to separate these frequencies by spectral and wavelet analysis.

\section*{Acknowledgments}

Authors are grateful to Prof. Vineamin Vityazev for useful advices
on the spectral
and wavelet analysis and T.~Shirai for fruitful discussion on FCN issues.

\end{document}